# Mist Computing: Principles, Trends and Future Direction

Manas Kumar Yogi [#1], K. Chandrasekhar[*2], G. Vijay Kumar[#3]

[#]Asst. Professor, Dept. Of CSE , Pragati College(Autonomous)
Surampalem,A.P. India

***Abstract*** — In this paper we present the novel idea of computing near the edge of IOT architecture which enhances the inherent efficiency while computing complex applications. This concept is termed as mist computing. We believe this computing will bring about an massive revolution in future computing technologies. instead of thrusting the control responsibility to gateways while data transmission the control is decentralised to end nodes which decrease the communicational delay of the network thereby increasing the throughput.

## I. INTRODUCTION (SIZE 10 & BOLD)

Scaling IOT to 75 billion devices is quite a handful of challenges. One way is to utilise the computing power at the edge of the network. Secondly, for the sake of minimising communication, develop measures to contain the computation at the edge of the network. Last but not the least solutions to scaling must e self managing ad self configuring. MIST computing helps in building large scale IOT systems. The IOT is regarded to have very small things at the very edge of the network like little power, limited RAM,ROM, limited communication bandwidth and not surprisingly may organisations refrain themselves from facing this challenges. The following table indicates the current utilization of bandwidth by IOT devices:

**TABLE I**

|  | CSR mesh (QUALCOMM) Single hop configuration | Zigbee Pro(si labs) 16 node network with high load | Thinnect mist 16 node network with high load |
|---|---|---|---|
| Average effective data rate (k/sec) | 0.017 | 0.05 | 0.1 |
| Duty cycle | 50% | 100% | 20% |

Even though the computing power is much more than what we had 10-15 years ago, the communication power is even much more than the computing power. We need 5 times more power to communicate with wireless devices. In case of a battery powered mesh network we have to communicate as little as possible to reduce the power consumption. For operation of IOT devices on the edge we need program memory size of 256kb and bandwidth of 250 Kbit/seconds. We infer easily that edge of IOT is not a scaled down version of the internet. So, while designing such a architecture we must account our needs before making the final design. The basic aim of mist computing is to rig computing to the very edge of the network ,i.e., sensors and actuators.IOT devices should not depend on internet as in real life ,physical systems won't be functional if there is a communication failure between the cloud ad the IOT device. The IOT devices should not have the capability to use the local intelligence using the guidelines that have been provided to act in case of a failure.

## II. GUIDING PRINCIPLES OF MIST COMPUTING

2.1 Network must provide information but not simply data.

2.2 The network should deliver only information that has been requested and only when it has been requested.

2.3 Dynamic creation of a system based on information needs with end devices working together using a subscriber provider model.

2.4 Devices must be situation aware ,they must adapt to the information needs and the network configuration. We should not have static bindings rules for device and data providers. The devices must dynamically discover the data providers and execute the application.

### A. Making Things Aware Of The Situation:

. The cloud and fog have awareness of the user needs and the global situation whereas the mist has awareness o the physical environment and the local situation, so together the responsibility is to execute an IOT application. In order to achieve this the global situation must be communicated to the edge devices and the edge devices must e ale to understand what does or how they need to behave in certain situations. The IOT application then actually spans from the very end of the edge network to the cloud. there are notable differences between edge computing ad mist computing. In edge computing, functionality is fixed, data processing occurs at the edge of the network, application configuration is fixed, whereas in mist computing, there are functionalities, timings which are dynamic and adjustable, there are high level application specific rules, new applications can e assembled from existing devices at runtime. For example, when we install a movement sensor the that movement it provides a service of movement detection, so this service can be used by any device which is then connected to that network. It can be used by a light sensor to turn the light on when there is movement or it can be used by an alarm system to provide an alarm if there should not e anyone in that area at that time. Performing all these operations using a 8-bit microcontroller and communication channel with provision for a packet of 100 bytes is much more difficult.

### III. DEVICE LEVEL RULES

1. Rules and data represented as 3 tuple similar to that of RDF triplets. These describe complex data relationships ad dependencies. A multi LISP type programming language is being developed which enables to communicate the commands across the network using a compact binary representation.

2. Its directly convertible to and from XML so its readable by semi humans because XML is both understandable by humans, machines.

*A. Routing in the Mist:*

Traditional wireless routing protocols are not suitable for routing in mist computing. here the routing protocols support device to device connectivity. it has been observed that sub-optimal routing paths increase the bandwidth requirements of the network. Also, in mist any node should be able to connect to any gateway eliminating dependence on a specific gateway. in a huge network, sometimes gateway failures force addition of a new gateway, the nodes near that gateway should have the ability to connect dynamically to such a node and a new route between itself and the gateway efficiently established so that the crisis of information unavailability is solved within the tolerance time.

### V. CONCLUSION

This paper serves as a concise effort in presenting the principles involved in mist computing technology. Although researchers are working appreciably towards developing this area of future computing it is still considered to be in nascent stage. Fog and Mist along

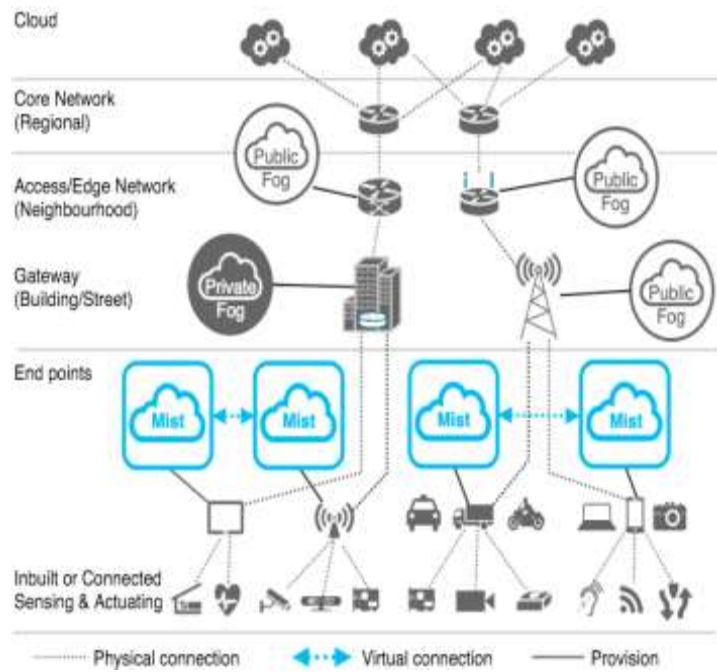

Fig. 1. Current Architecture Of Mist Computing

As we can observe in above representation, the mist nodes are responsible to process the data which has to e handed over to the IOT devices which include the sensors and actuators with a physical connection among themselves. the mist node as also monitor the quality of link parameters. At the gateway level the functionalities are loading updated application rules and tuning application parameters, monitoring the health of local nodes ,execution of computationally intensive services .finally at the cloud level new applications can be deployed and applications can be coordinated along with service quality monitoring and monitoring health of the running applications.

### IV. APPLICATION IN REAL LIFE

The company named Thinnect under the leadership of the Jugo Preden, CEO of Thinnect has successfully implemented a mist computing stack from the physical contains a robust rule based application logic framework. The memory requirement of the entire stack is about 100kbytes.it has been deployed and tested by DEFENDEC for 6 years in a security application and recently in a smartcity application in many countries by a organisation named CITYNTEL.

with cloud would be beneficial for the entire distributed computing ecosystem if all the components were able to provide more functionality and more capable microcontrollers for the network nodes. Fog and mist computing are complementary to each other . The application jobs, which are comparatively more computationally intensive can be executed in the gateway whereas the one which is less computationally intensive tasks can be executed in the end devices. As an application in Mist computing, it is treated as a collection of hierarchical services which can be distributed among the computing nodes liberally.